\documentclass[twoside,11pt]{article}
\newcommand{\TODO}[1]{}

\newcommand{\R}{\mathbb{R}}
\newcommand{\C}{\mathbb{C}}

\newcommand\Cl{\mathfrak{Cl}}
\newcommand\Id{\mathsf 1}
\newcommand{\Dir}{\mathcal{D}}
\newcommand{\pd}[1]{\frac{\partial}{\partial{#1}}}
\newcommand\eq[1]{{\rm Eq.~(\ref{#1})}}
\newcommand\Sec[1]{Sec.~\ref{Sec:#1}}
\newcommand\mPsi{\mathsf{\Psi}}
\newcommand{\LA}{\mathfrak S}
\newcommand{\Ext}{\Lambda}
\newcommand{\Lin}{\mathcal{L}}

\newcommand{\KG}{\square}

\newcommand{\nsum}{\sum\nolimits}
\newcommand{\Mat}[2][rrrr]{\left(\begin{array}{#1}#2\end{array}\right)} 

\newcommand{\DirHod}{\check\Dir}
\newcommand{\ael}[1]{\mathtt{#1}}
\newcommand{\e}{\mathsf e}
\newcommand{\Alt}{\mathcal A}
\newcommand{\ClEx}{\mathfrak{E}}
\newcommand{\ClExm}{\mathfrak{E}_\mathfrak{m}}
\newcommand{\M}{\mathcal M}

\newcommand{\fnorm}[2]{\mathopen{<}#1\mathbin{,}#2\mathclose{>}}

\def\paper{article}

\makeatletter
 \@addtoreset{equation}{section} 
 \renewcommand{\theequation}{\thesection.\@arabic\c@equation} 
\makeatother

\usepackage{amsmath,amssymb,exscale}
\usepackage{graphics}

\title{The Dirac Equation and General Linear Transformations of Coordinate Systems}
\date{}

\author{{\em Alexander Yu.\ Vlasov}
}
\begin{document}
\maketitle
\pagestyle{myheadings} 
\thispagestyle{plain}         
\markboth{Alexander Vlasov}{Dirac Equation and General Linear Transformations} 
\setcounter{page}{1} 

{\centering \includegraphics{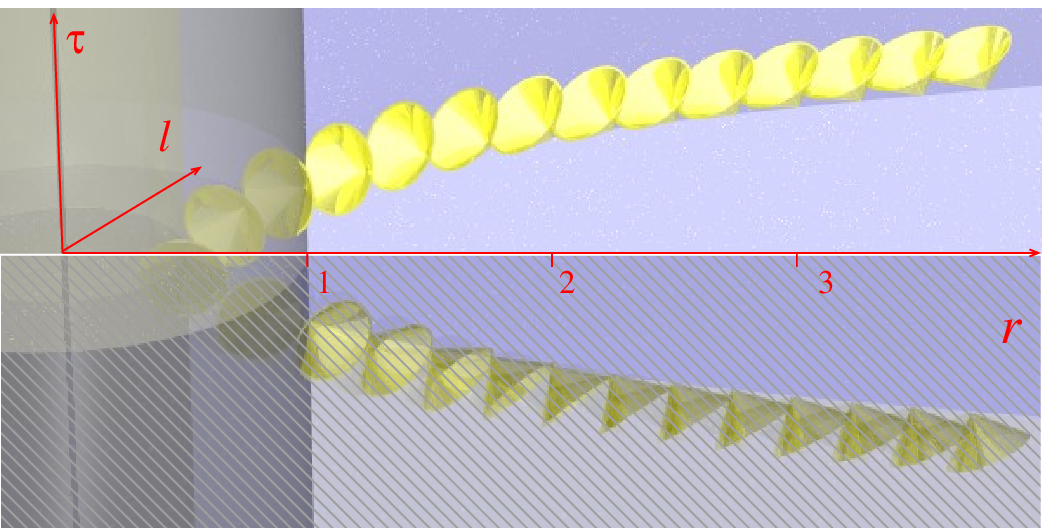}

}
\begin{abstract}
The spinor representation of the Lorentz group does not accept simple generalization
with the group $GL(4,\R)$ of general linear coordinate transformations. The Dirac 
equation may be written for an arbitrary choice of a coordinate system and a metric,
but the covariant linear transformations of the four-component Dirac spinor
exist only for isometries. For usual diagonal Minkowski metric the
isometry is the Lorentz transformation. On the other hand, it is possible to
define the Dirac operator on the space of anti-symmetric (exterior) forms, and
in such a case the equation is covariant for an arbitrary general linear
transformation. The space of the exterior forms is sixteen-dimensional,
but usual Dirac equation is defined for four-dimensional complex space
of Dirac spinors. Using suggested analogy, in present paper is discussed
possibility to consider the space of Dirac spinors as some ``subsystem'' 
of a bigger space, where the group $GL(4,\R)$ of General Relativity acts 
in a covariant way.

  For such purposes in this \paper\ is considered both Grassmann algebra
of complex anti-symmetric forms and Clifford algebra of Dirac matrices.
Both algebras have same dimension as linear spaces, but different structure
of multiplication. The underlying sixteen-dimensional linear space also may be 
considered either as space of complex $4\times 4$ matrices, or as space of states 
of two particles: the initial Dirac spinor and some auxiliary system. 
It is shown also, that such approach is in good agreement with well known idea
to consider Dirac spinor as some ideal of Clifford algebra. Some other
possible implications of given model are also discussed.
\end{abstract}

\section{Introduction}
\label{Sec:Intro}
\subsection{Klein-Gordon and Dirac Equations}
Let us consider an operator of second order
\begin{equation}
 \KG = -\nsum_{kj} g_{kj}\pd{x_k}\pd{x_j}.
\label{dAl}
\end{equation}
For Minkowski metric $g$ of (flat) spacetime the operator \eq{dAl} is the
d'Alembert (wave) operator and components $\psi$ of a relativistic field 
for a particle with mass $m$ satisfy to a Klein--Gordon equation 
(in the system of units $\hbar=1$, $c=1$). 
\begin{equation}
 (\KG - m^2)\psi = 0.
\label{KGE}
\end{equation}

For a relativistic particle with spin $1/2$, the Klein-Gordon
operator should be rewritten as \cite{QFT}
\begin{equation}
 \KG = \Dir^2, \quad \KG - m^2 = (\Dir - m)(\Dir + m),
\label{fact}
\end{equation}
where 
\begin{equation}
 \Dir = \sum_{k=0}^3 i \gamma^k \pd{x_k}
\end{equation}
is Dirac operator and $\gamma_k$ are four $4 \times 4$ complex matrices
with property
\begin{equation}
 \gamma^j \gamma^k + \gamma^k \gamma^j = 2 g_{kj} \Id,
\label{gcom}
\end{equation}
where $\Id$ is unit matrix.

The Dirac equation for a four-component spinor $\psi \in \C^4$ may be written as
\begin{equation}
 \Dir \psi = m \psi,
 \quad \psi \in \C^4,
 \quad \Dir = \sum_{k=0}^3 i \gamma^k \pd{x_k}
\label{DirEq}
\end{equation}
and due to \eq{fact} any component of $\psi$ also satisfies to the Klein-Gordon
equation \eq{KGE}.

Really the factorization $\Dir^2 = \KG$ used in \eq{fact} together
with \eq{gcom} may be written for any metric $g$ (and dimension) \cite{ClDir}.

\subsection{Two-component Spinors}

It is also convenient to consider construction of the four-component Dirac spinors
as a pair of two-component (Pauli\footnote{The two-component spinors also are called 
``Weyl spinors,'' ``half-spinors'' or simply ``spinors'' --- in the last case the 
four-component spinors are called ``bispinors.''}) spinors \cite{DauIV}. Here we are
again considering factorization of the Klein--Gordon equation, but instead of the 
formal decomposition \eq{fact}, it is possible to write two equations for Pauli 
spinors $\eta,\xi \in \C^2$
\begin{equation}
 i(\pd{x_0}+\sum_{k=1}^3\sigma_k\pd{x_k})\eta = m \xi,
\quad
 i(\pd{x_0}-\sum_{k=1}^3\sigma_k\pd{x_k})\xi = m \eta
\label{Pauli2}
\end{equation}
with $2 \times 2$ Pauli matrices 
\begin{equation}
\sigma_1 \equiv \sigma_x = \Mat{0&1\\1&0}, \quad
\sigma_2 \equiv \sigma_y = \Mat{0&-i\\i&0}, \quad
\sigma_3 \equiv \sigma_z = \Mat{1&0\\0&-1}.
\label{PauliMat}
\end{equation}

From such a point of view, the four-component Dirac spinor may be considered
as a composition of two Pauli spinors
\begin{equation}
\psi = \Mat{\eta \\ \xi}
\end{equation}
and then two equations \eq{Pauli2} are equivalent to one Dirac equation
\eq{DirEq} with $4 \times 4$ matrices \cite{DauIV}
\begin{equation}
 \gamma^0 = \Mat{0 & 1 \\ 1 & 0}, \quad
 \gamma^k = \Mat[cc]{0 & -\sigma_k \\ \sigma_k & 0} \quad (k = 1,2,3)
\label{GamSpin}
\end{equation}
(where $0$ and $1$ are also $2 \times 2$ matrices). Of course, the matrices
\eq{GamSpin} are satisfying \eq{gcom} for Minkowski metric.

\subsection{Dirac--K\"ahler Equation}

Factorization of d'Alembertian \eq{fact} using Dirac operator, has some 
resemblance with decomposition of Laplace--Beltrami operator in the Hodge theory 
\cite{Warner}
\begin{equation}
 \Delta = d d^\star + d^\star d = -(d - d^\star)^2,\quad 
 d^\star = \pm \star d \star, \quad
 d^2 = {d^\star}^2 = 0 
\label{Hodge}
\end{equation}
and it is possible to introduce representation of the Dirac operator on 
space of anti-symmetric (exterior) forms \cite{ClDir,Geom4}. In such a case an 
analogue of the Dirac equation should be written as
\begin{equation}
 \DirHod \Upsilon = m \Upsilon, \quad \DirHod \equiv (d - d^\star)
\label{DirForm}
\end{equation}
where $\Upsilon \in \Ext$ is some anti-symmetric form. Such a kind of Dirac 
equation is widely used since introduction by Landau and Ivanenko at 1928
and K\"ahler at 1962 \cite{Yura,Marchuk,DirHes}. 
It should be mentioned, that because $d: \Ext^p \to \Ext^{p+1}$
and $d^\star: \Ext^p \to \Ext^{p-1}$, the Dirac--K\"ahler operator $d - d^\star$ 
does not respect the order $p$ of a form, unlike the Laplace--Beltrami operator 
$\Delta: \Ext^p \to \Ext^p$.

\subsection{Covariance}

A multi-component quantum field should be
{\em covariant} with respect to coordinate transformations \cite{QFT}, 
{\em e.g.}, components of the quantum field $u(x)$ after change of coordinate 
system $L: x \to x'$ 
also are changed $u' = \LA_L u$, where $\LA_L$ is some representation of
group of coordinate transformations acting on the space of multi-component
wave vectors.

Different groups of acceptable coordinate transformations may cause particular
limitations to a model. Say, the two-component spinors discussed 
above are covariant not only with respect to group $SO(3)$ of 
transformations of non-relativistic space (3D rotation), but also with respect 
to Lorentz group $SO(3,1)$. 
On the other hand, the Pauli spinors are not covariant with respect
to a bigger group $O(3,1)$ with the time reflection and it may be considered as
a reason of the necessity to consider the four-component Dirac spinors \cite{DauIV}. 
  
The Dirac spinors are not covariant with respect to group $GL(4,\R)$ of all possible 
coordinate transformations. 
On the other hand, the anti-symmetric forms in \eq{DirForm} and operators 
$d, d^\star$ used for construction \eq{Hodge} of the Dirac--K\"ahler operator are 
covariant with respect to the general linear group $GL(4,\R)$ of coordinate 
transformations, it is well known property of the differentials and tensor fields. 
 
The transition from space of two-component spinors to the bigger space of 
four-component spinors let us consider covariance with respect to $O(3,1)$ 
instead of $SO(3,1)$ and $O(3)$. Formally, the transition from space of 
four-component spinors to the bigger sixteen-dimensional space of 
anti-symmetric forms  made the Dirac equation covariant with respect to the 
group $GL(4,\R)$ of all coordinate transformations. 

It is possible also to consider a space of exterior forms with complex 
coefficients and use other representations with sixteen-dimensional
linear spaces, for example the space of $4 \times 4$ matrices, or tensor 
product of two four-dimensional spaces. The last example also 
formally corresponds to consideration of a compound quantum system 
$\C^4 \otimes \C^4$ with two particles described by the Dirac spinors. 
In such a case relation with the initial Dirac equation is not so simple, 
as in the example with transition between two- and four-component spinors. 

\subsection{Outline}

The present \paper\ is devoted to consideration of such $GL(4,\R)$-covariant 
extension of Dirac equation with a sixteen-dimensional linear space, together 
with description of relation with usual Dirac equation for four-components 
spinors and some implications of such generalization.
  
It is extension and development of an earlier work \cite{GrDir03}. 
In \Sec{Clif} and \Sec{Spin} are described necessary methods from
theory of Clifford algebras and Spin groups. Exterior forms and Hodge
theory are revisited in \Sec{Ext}. Transformation properties of
Dirac equation are discussed in \Sec{DirTr}. In \Sec{GR} are discussed
some applications of group $GL(4,\R)$ of general linear coordinate 
transformations to General Relativity, because it is useful for 
justification of suggested approach with generalization of Dirac equation.

\TODO{... more details, about THAT it is ...,
       more detailed plan: Sec 2, Sec 3, ...}

\section{Clifford Algebra of a Quadratic Form}
\label{Sec:Clif}
\subsection{Definition of Clifford Algebra}

The decomposition of the wave operator \eq{dAl} in \Sec{Intro} follows to a 
standard procedure of representation of a quadratic form as the square of an 
element of some algebra \cite{ClDir,Post}. 
If there is $n$-dimensional vector space $V$ with a quadratic form 
$g(v) \equiv g(v,v)$, $v \in V$, some algebra $\Cl$ and $n$ elements (generators)
$\e_k \in \Cl$, it is possible to consider the linear embedding $\iota : V \to \Cl$,
$\iota(v) = \sum v_k \e_k \in \Cl$, where $v_k$ are components of the vector $v$ in 
some basis. Let us denote $\iota(v)$ simply as $\ael v$.
In such a case it is reasonable to look for expression of the quadratic form $g$ as 
the square of $\ael v \in \Cl$
\begin{equation}
 -g(v) \Id = \ael v^2,\quad \ael v \equiv \iota(v),
\label{ClifSq}
\end{equation}
where $\Id$ is the unit of the algebra, and the minus sign is due to some traditions 
\cite{ClDir} (``hypercomplex numbers''), but may be omitted in some accounts 
\cite{Post}. It is simply to check, that from \eq{ClifSq} directly follows an 
equivalent of \eq{gcom}
\begin{equation}
 \{\e_k,\e_j\} \equiv \e_k \e_j + \e_j \e_k = -2 g_{kj} \Id.
\label{ecomm}
\end{equation}
Due to \eq{ecomm} $n$ generators $\e_k$ may produce up to $2^n$ linearly 
independent products and it is the maximum possible dimension of the Clifford 
algebra generated by these elements \cite{ClDir}.

Subspaces generated by products of $k$ different generators are denoted here 
$\Cl^{(k)}$
\begin{equation}
\Cl^{(k)} = \mathrm{span}\{\e_{i_1} \cdots \e_{i_k} : 
1 \le i_1 < i_2 < \cdots < i_k \le n\}, 
\quad \dim \Cl^{(k)} = \binom{n}{k},
\label{Clk}
\end{equation}
and are called sometime {\em $k$-multivectors}. It is possible to decompose
the Clifford algebra in the direct sum of $n+1$ such subspaces 
$\Cl = \Cl^{(0)}\oplus \Cl^{(1)} \oplus \cdots\oplus  \Cl^{(n)}$.

\subsection{Grassmann Algebra}

A particular case \cite{ClDir} is the trivial, degenerate quadratic form 
$g(v)\equiv 0$, $\forall v \in V$. For such a case \eq{ecomm} correspond to
the definition of {\em the Grassmann algebra}. In such a case product of
elements $\ael a \ael b$ often denoted $\ael a \wedge \ael b$ and
instead of \eq{ClifSq} and \eq{ecomm} we have
\begin{equation}
\ael v^2 = \ael v \wedge \ael v = 0; \qquad
 \e_k \wedge \e_j + \e_j \wedge \e_k = 0 \quad (\forall j,k).
\label{acom}
\end{equation}
The Grassmann algebras devote separate consideration, see \Sec{Ext}.
For any quadratic form $g$ it is possible to use decomposition 
$V = R \oplus R^\perp$, with degenerate and nondegenerate subspaces $R$ and 
$R^\perp$ respectively \cite{ClDir}, and so it is reasonable to describe
classification of the Clifford algebras with nondegenerate quadratic forms.

Sometime it is also convenient for complex Clifford algebras with even dimension 
and the diagonal form (like \eq{eucom} below) to consider new set of generators
\begin{equation}
 \ael a_j = \frac{1}{2}(\e_j + i \e_{m+j}),
\quad
 \ael a_j^* = \frac{1}{2}(\e_j - i \e_{m+j}),
\quad j = 1,\ldots,m .
\label{adef}
\end{equation}
Then both subsets with $m$ elements $\ael a_j$ or $\ael a_j^*$ generate Grassmann 
subalgebras of the initial Clifford algebra. Together with usual anticommutation
\eq{acom} of all elements in each algebra, there is specific relations between pair
of elements from different subsets, {\em i.e.}
\begin{equation}
 \{\ael a_j,\ael a_k\} = \{\ael a_j^*,\ael a_k^*\} = 0,
\quad
 \{\ael a_j,\ael a^*_k\} = 2 \delta_{jk}.
\label{CAR}
\end{equation}
The \eq{CAR} coincide with canonical anticommutation relations (CAR) for annihilation
and creation operators for fermions. Choice of representation of $\e_j$ with tensor 
product of Pauli matrices (see \eq{defE} below) corresponds to the usual 
Jordan--Wigner formalism \cite{WeylGQM}
\begin{eqnarray}
 \ael a_{k} & = &
  {\underbrace{\Id\otimes\cdots\otimes \Id}_{m-k-1}\,}\otimes
 a\otimes\underbrace{\sigma_z\otimes\cdots\otimes\sigma_z}_k \, ,
 \nonumber\\
 \ael a_{k}^* & = &
 {\underbrace{\Id\otimes\cdots\otimes \Id}_{m-k-1}\,}\otimes
 a^+\otimes\underbrace{\sigma_z\otimes\cdots\otimes\sigma_z}_k \, ,
 \label{JorWig}
\end{eqnarray}
where $\Id$ is $2 \times 2$ unit matrix and
\begin{equation}
 a = \Mat{0&1\\0&0}, \quad a^+ = \Mat{0&0\\1&0}.
\end{equation}
So we have two different representations of the real Grassmann algebra with $m$ 
generators in the same algebra of $2^{m} \times 2^{m}$ real matrices.

\subsection{Nondegenerate Quadratic Forms}

For a nondegenerate quadratic form it is always possible to choose {\em the 
normalized basis}. In such a basis the matrix of the quadratic form is diagonal 
$g_{jk} = \pm \delta_{jk}$ and it is convenient to use such forms
for classifications of {\em the real} Clifford algebras \cite{ClDir,Post,Port}.
Here the algebra for the quadratic form with $l$ positive and $m$ negative terms
is denoted as $\Cl(l,m)$. For a complex vector space the signs do not matters,
because they always may be changed using multiplication on $i$ and so for Clifford
algebra with complex coefficients here is used notation $\Cl(n,\C)$.

For the complex case classification of the Clifford algebras $\Cl(n,\C)$ is simpler.
For even $n=2m$ there is an isomorphism with the algebra of $2^m \times 2^m$
complex matrices and for odd case $2m+1$ with the direct sum of two matrix 
algebras \cite{ClDir}
\begin{equation}
 \Cl(2m,\C) \cong M(2^m,\C);
\quad
  \Cl(2m+1,\C) \cong M(2^m,\C) \oplus M(2^m,\C).
\label{CxClCls}
\end{equation}

The generators $\e_k$ of the Clifford algebra for even dimension $\Cl(2m,\C)$
satisfying \eq{ecomm} for Euclidean form
\begin{equation}
 \{\e_k,\e_j\} \equiv \e_k \e_j + \e_j \e_k = 2 \delta_{kj} \Id.
\label{eucom}
\end{equation}
may be expressed using Pauli matrices as \cite{ClDir,WeylGQM}
\begin{eqnarray}
 \e_{2k} & = &
  {\underbrace{\Id\otimes\cdots\otimes \Id}_{m-k-1}\,}\otimes
 \sigma_x\otimes\underbrace{\sigma_z\otimes\cdots\otimes\sigma_z}_k \, ,
 \nonumber\\
 \e_{2k+1} & = &
 {\underbrace{\Id\otimes\cdots\otimes \Id}_{m-k-1}\,}\otimes
 \sigma_y\otimes\underbrace{\sigma_z\otimes\cdots\otimes\sigma_z}_k \, ,
 \label{defE}
\end{eqnarray}
where $k = 0,\ldots,m-1$ and $\Id$ is $2\times 2$ unit matrix.

The algebra $\Cl(2m+1,\C)$ may be considered as a subspace of $\Cl(2m+2,\C)$ with 
a reduced set of generators \eq{defE} (without $\e_{2m+2}$). It is also convenient 
to change $\sigma_x$ to $\sigma_z$ in the last generator 
\begin{equation}
 \e_{2m+1} =  \underbrace{\sigma_z\otimes\cdots\otimes\sigma_z}_{2m+1} \, ,
\end{equation}
because in such a case the structure $\Cl(2m+1,\C) \cong \Cl(2m,\C) \oplus \Cl(2m,\C)$
is more transparent.

For real Clifford algebras $\Cl(l,m)$ the classification is more difficult.
Different cases may be arranged using a $8 \times 8$ table ($l,m = 0,\ldots,7$)
and recurrent structure of $\Cl(8k+l, 8j+m)$ \cite{Port} instead of only 
two different structures $\Cl(2k+l)$, $l = 0,1$ in complex case. Anyway, all the real 
Clifford algebras may be represented either as a some matrix algebra with real, 
complex, quaternion coefficients, or direct sum of two such algebras. 

\medskip

{\em The space of spinors --- is a space of representations of the Clifford algebras 
(and Spin groups, see \Sec{Spin} below)}. 
So, if a Clifford algebra is isomorphic with algebra of 
$N \times N$ (real, complex, quaternion) matrices, the spinors are 
$N$-dimensional (real, complex, quaternion\footnote{Quaternions are
not commutative, so formally instead of linear space should be used 
left quaternion module.}) linear spaces.

It should be mentioned, that $\Cl(l,m)$ for a particular choice of $l$ and $m$
may contain an element $\ael i \in \Cl$ commuting with all other elements of
algebra \TODO{more details} and with property $\ael i^2 = -\Id$. Formally it should be
distinguished from a complex Clifford algebra, because for construction
was used the real linear space and so $\ael i$ here is not a number (``scalar''),
but an element of the real algebra. For example
\begin{equation}
\Cl(3,0) \cong \Cl(1,2) \cong M(2,\C),\quad
\Cl(4,1) \cong \Cl(2,3) \cong M(4,\C),
\end{equation}
are real Clifford algebras (see \cite{Port}) and 
\begin{equation}
\Cl(2,\C) \cong M(2,\C),\quad
\Cl(4,\C) \cong M(4,\C)
\end{equation}
are complex Clifford algebras.

\subsection{Transformation Properties}
\label{Sec:ClifTrans}

Let us suggest, that we have built a Clifford algebra for some quadratic form 
$g$ and want to rewrite the relations \eq{ecomm} for a new set of elements
\begin{equation}
 \e'_k = \nsum_j A_{kj} \e_j.
\label{Ae}
\end{equation}
Then the set of equations \eq{ecomm} for the new elements is
\begin{equation}
-2g'_{kl} = \{\e'_k,\e'_l\} = \sum_{jm}\{A_{kj}\e_j,A_{lm}\e_m\} =
-2 \sum_{jm}A_{kj}A_{lm}g_{jm},
\end{equation}
and using a matrix notation last equation may be rewritten as
\begin{equation}
 g' = A g A^T,
\label{AgAT}
\end{equation}
where $A$ is the matrix of coefficients $A_{kj}$ of transformation \eq{Ae} 
and $A^T$ is the transposed matrix. 
The \eq{AgAT} coincides with the formula for change of the metric after
transition (described by the matrix $A$) to other basis in the space $V$. \TODO{? inverse}

So ``algebraic square root'' \eq{ClifSq} of an arbitrary (not necessary diagonal) 
quadratic form $g'$ may be found using diagonalization of given form and equation 
\eq{AgAT}. It is necessary first to convert the quadratic form to a sum of squares
using appropriate transformation of a basis in $V$ and to find a Clifford algebra 
for the diagonal form. Next, with $n$ generators $\e_k$ of given algebra, it is
possible using \eq{Ae} to build the a set with $n$ elements $\e'_j$ satisfying
\eq{ecomm} for the initial quadratic form $g'$.

Such a technical way to describe a construction of the spinors formally includes
few objects: the vector space $V$ with the metric $g$, a diagonal metric $g_0$ and
a matrix $A$ (``vielbein'') of transition from normalized form $g(v,w) = g_0(Av,Aw)$.  
It is relevant to {\em a tetrad (vierbein) formalism of the General Relativity} 
(see \Sec{SpinLor} and references therein).

Really the tetrad formalism does not necessary suggest the choice of $g_0$ in
a diagonal form, it may be any other form convenient for description of the 
Clifford algebra. For example, it is possible to choose ``an isotropic basis'' 
\eq{adef}. 
The main property of a tetrad representation --- is using some fixed form $g_0$
and necessity of ``vielbein'' $A$ for transition to concrete metric $g$.  
An example with isotropic tetrads is Newman--Penrose formalism \cite{SpinI,MTBH}.

On the other hand, for complete description of a Clifford algebra $\Cl$ with a
quadratic form $g$, it is enough to give only a vector space $V$ with the metric $g$,
the algebra $\Cl$ and the embedding $\iota : V \to \Cl$, $g(x) = -\iota^2(x)$ used in 
the definition \eq{ClifSq}. In such a case, $A$ and $g_0$ look as redundant objects.
Covariant character of such definition may be more clear from construction
with exterior (Grassmann) algebra discussed below in \Sec{Ext}.

\subsection{Automorphisms}

It should be mentioned, that \eq{Ae} might not be considered as some transformation
of the {\em whole} Clifford algebra. The matrix $A$ describes transformation of the 
vector space $V$ and an $n$-dimensional subspace of the algebra defined as embedding 
$\iota(V)$ or as the linear span of the elements $\e_k$, but it is not possible 
without additional suggestions to apply it to an arbitrary element of the algebra 
like $\e_j\e_k$, $\e_j\e_k\e_l$, {\em etc}.

One method to extend such transformations from the generators to the whole algebra, 
is to use {\em automorphisms}, {\em i.e.} the linear maps  to itself 
$\alpha : \Cl \to \Cl$ with property
\begin{equation}
 \alpha(\ael a\,\ael b) = \alpha (\ael a)\,\alpha(\ael b)
\label{AutAlg}
\end{equation}
for any elements $\ael{a,b}$ of the algebra.

A particular example is {\em the internal automorphism}, defined as
\begin{equation}
 \alpha_{\ael h}(\ael a) = \ael h\,\ael a\,\ael h^{-1},
\end{equation}
where $\ael h$ is an arbitrary invertible element of the algebra.

Due to \eq{ecomm} any automorphism of a Clifford algebra saves the same quadratic 
form for the new set of generators $\tilde\e_k = \alpha(\e_k)$
\begin{equation}
  \{\tilde\e_k,\tilde\e_j\} = \alpha(\e_k)\alpha(\e_j)+\alpha(\e_j)\alpha(\e_k) =
 \alpha(\e_k\e_j+\e_j\e_k)=-2g_{kj}\alpha(\Id)=-2g_{kj}\Id.
\end{equation}

\medskip

So, it is an important question, if the map \eq{Ae} corresponds to some automorphism
of the algebra. It is useful to consider two different examples. 

\begin{itemize}
\item[{\bf 1.}] 
It was shown that the Grassmann algebra is equivalent to a Clifford 
algebra with a degenerated quadratic form $g \equiv 0$ \eq{acom}. Any 
coordinate transformations saves such a trivial form and for any $A$ \eq{Ae} 
may be extended to an automorphism of the whole algebra, it corresponds to usual
covariant transformations of anti-symmetric tensors $\Ext = \bigoplus \Ext^k$
with ranges $0 \le k \le n$ ({\em e.g.} see \eq{TensTransf} in \Sec{Ext}). 
The automorphism may not be internal, because
all elements of Grassmann algebra, except $\Id$, do not have inverse.

\item[{\bf 2.}] 
For the Clifford algebra a with nondegenerate form $g$ only {\em isometries}
$A$ saves the quadratic form (by definition). So only for isometries \eq{Ae}
could be expressed as some automorphism, and really always exists {\em an internal
isomorphism} with necessary property, {\em i.e.} for any isometry $A$
exists element $\ael h_A \in \Cl$:
\begin{equation}
 \iota(A v) = \ael h_A\, \iota(v)\, \ael h_A^{-1} \quad (v \in V),
\label{ihah}
\end{equation}
{\em i.e.}, instead of \eq{Ae} for isometry $A$ it is possible to
write
\begin{equation}
 \ael e'_j = \ael h_A\,\ael e_j\,\ael h_A^{-1} 
 \quad (\forall j, \mbox{ no summation}),
\label{hah}
\end{equation}
but such element $\ael h_A$ exists if and only if $g(Av)=g(v)$. 
It corresponds to construction of {\em Spin group} for given Clifford 
algebra \cite{ClDir,Post}.
\end{itemize}

\TODO{Principal (anti-)automorphism}

\section{Spin Groups}   
\label{Sec:Spin}
\subsection{Reflections and Rotations}

A simple example of the isometry is {\em the reflection}, $R_v$, defined by
some $v \in V$, $g(v) \ne 0$ as
\begin{equation}
 R_v: x \mapsto x - 2 \frac{g(x,v)}{g(v)} v.
\label{Refl}  
\end{equation}
Any isometry of $n$-dimensional space may be written as composition of 
few reflections \cite{ClDir,Cart}.
 
\begin{figure}[ht]
\begin{center}
\includegraphics{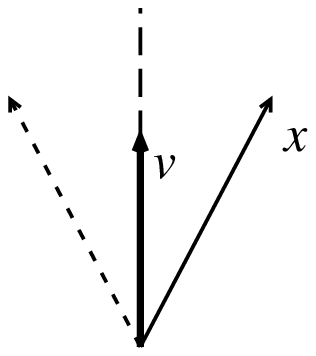}~
\includegraphics{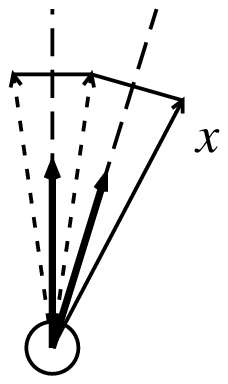}
\caption{Example of reflection and presentation of rotation
 as two reflections for two-dimensional plane}\label{Fig:reflrot}
\end{center}
\end{figure}

It may be checked directly \cite{ClDir}, that the reflection may be simply
written by using representation of the vector as an element of the Clifford algebra
\begin{equation}
 R_v: \ael x \mapsto \ael v\, \ael x\,\ael v^{-1}.
\label{ClRefl} 
\end{equation}
It is also possible to consider only elements with $g(v)=\pm 1$, because
for any number $\lambda \ne 0$, the vectors $v$ and $\lambda v$ represent 
the same reflection.

\subsection{Spin, Pin and Spoin Groups}
\label{Sec:SpinPin}

In both Euclidean and pseudo-Euclidean case, any element of the group 
$SO(l,m)$ may be represented as a composition of {\em even} number of 
reflections, and so, if to consider {\em the group $Spin(l,m)$, consisting of all 
possible even products} $\ael h  = \ael v_1 \cdots \ael v_{2k} \in \Cl(l,m)$,
for $v_j \in V$, $g(v_j)=\pm 1$, we have representation of the group $SO(l,m)$ as 
\begin{equation}
 \ael x \mapsto \ael h \, \ael x \, \ael h^{-1}, 
 \quad \ael h \in Spin(l,m)
\label{SpinAct}
\end{equation}
and because $\ael h$ and $-\ael h$ correspond to the same transformation,
it is $2 \to 1$ representation of $SO(l,m)$ group. Similarly, {\em all
products with an arbitrary (not necessary even) number of elements $\ael v_j$
produce group} $Pin(l,m)$, it is $2{\to}1$ representation of $O(l,m)$ group
\cite{ClDir,Post}. 

Let us consider {\em the linear span of all possible products with even 
number of generators}. The subspace is closed under multiplication, includes 
unit and so may be considered as {\em a new Clifford algebra}, generated by 
$n-1$ generators $\acute\e_j = \e_j\e_n$, $j=1,\ldots,n-1$.
Spin group also may be considered using this Clifford algebra with 
smaller dimension. Such ``economical'' representation of Spin group
sometime called {\em the Spoin group}, but in such a case, instead of
internal automorphism \eq{SpinAct} should be used changed expression 
\begin{equation}
 \ael x \mapsto \ael h\, \ael x\, \ael h'^{-1}, 
 \quad \ael h \in Spoin(l,m)
\label{SpoinAct}
\end{equation}
where $\ael h'$ is specific ``Clifford conjugation'' \cite{ClDir}.

The relation of Spoin and Spin groups is important for discussion about
representations of the Lorentz group with Pauli and Dirac matrices 
in \Sec{SpinLor}.

\section{Grassmann Algebra of Exterior Forms}
\label{Sec:Ext}
\subsection{Forms and Tensors}

Let us recall some standard definitions and properties \cite{Warner,Geom4,KobNomI,MTBH}.
The {\em tensor of type $(r,s)$} may be written using a basis
$e_j$ in an $n$-dimensional vector space $V$ and $e^j$ in the dual space $V^*$
(linear functions on $V$) as 
\begin{equation}
 T = \sum T^{i_1\ldots i_r}_{j_1\ldots j_s} 
 e_{i_1} \otimes \cdots \otimes e_{i_r} \otimes e^{j_1} \otimes \cdots \otimes e^{j_s}.
\end{equation}
For a change of the basis described
by the matrix $A$, $e_i = \sum_j A_i^j e'_j$ such a tensor is transformed as 
\begin{equation}
 T'^{i_1\ldots i_r}_{j_1\ldots j_s} = 
\sum A^{i_1}_{k_1}\cdots A^{i_r}_{k_r}B^{m_1}_{j_1}\cdots B^{m_s}_{j_s}%
T^{k_1\ldots k_r}_{m_1\ldots m_s}
\label{TensTransf} 
\end{equation}
where $B=A^{-1}$ is the inverse of the matrix $A$ \cite{KobNomI}.
\TODO{To explain, why in opposite directrion (?)} 

An exterior $s$-form $\omega \in \Ext^s(V)$ is {\em the totally antisymmetric tensor} 
of type $(0,s)$. Any tensor of type $(0,s)$ may be considered as multilinear functional
and $s$-form is a specific one, antisymmetric in every pair of arguments
\begin{equation}
 \omega(v_1,\ldots,v_i,\ldots,v_j,\ldots,v_s) =
 -\omega(v_1,\ldots,v_j,\ldots,v_i,\ldots,v_s), \quad \forall i,j. 
\end{equation}
Let us consider an operation of {\em alternation} of a tensor of type $(0,s)$
\begin{equation}
\Alt T(v_1,\cdots,v_s) 
 = \frac{1}{s!}\sum_\pi (-)^\pi T(v_{\pi(1)},\cdots,v_{\pi(s)}),
\end{equation}
where $\pi$ denotes all $s!$ permutations of $s$ indexes and $(-)^\pi$ is
$+1$ for an even permutation and $-1$ for an odd one,
then $\Alt T$ is always antisymmetric tensor. 
The basis of space of the antisymmetric tensors may be written using an exterior 
(wedge) product 
\begin{equation}
 e^{j_1} \wedge \cdots \wedge e^{j_s} = \Alt(e^{j_1} \otimes \cdots \otimes e^{j_s}).
\end{equation}
For arbitrary forms $\omega \in \Ext^r$, $\vartheta \in \Ext^s$ it may be defined
\begin{equation}
\omega \wedge \vartheta \equiv \Alt(\omega \otimes \vartheta),\quad
\omega \wedge \vartheta = (-1)^{rs}\vartheta \wedge \omega.    
\end{equation}

A maximal possible range of the exterior form is $s=n$ and the space of all exterior 
forms is denoted as
\begin{equation}
 \Ext(V) = \bigoplus_{j=0}^n \Ext^j(V),\quad \dim \Ext(V) = 2^n,
 \quad \dim \Ext^j(V) = \binom{n}{j}.
\end{equation}

It is possible to consider the tensor fields on an $n$-dimensional manifold. For 
an exterior form $\omega$ it is also possible to introduce {\em exterior 
differentiation} $d\omega$, a linear operator from space of $s$-forms to $(s+1)$-forms
$d : \Ext^{s}(V) \to \Ext^{s+1}(V)$ with properties
\begin{enumerate}
\item For function, {\em i.e.} $f \in \Ext^0(V)$, it is usual {\em total differential}.
\item For $\omega \in \Ext^r(V)$ and $\vartheta \in \Ext^s(V)$:
 $d(\omega \wedge \vartheta) = 
 (d\omega) \wedge \vartheta + (-1)^r \omega \wedge (d\vartheta)$.
\item ``Poincar\'e $d$-lemma:'' $d^2 = 0$.
\end{enumerate}
\TODO{... tangent space ..., write formula for d directly}

\subsection{Hodge Theory}
\label{Sec:Hodge}

Let us consider spaces with scalar product (metric). If some metric $\fnorm\cdot\cdot$ 
is introduced in $\Ext^1(V) \cong V^*$, it is possible to extend it on whole $\Ext(V)$.
For two forms with different ranges it is zero, otherwise the norm \cite{Warner}
\begin{equation}
\fnorm{w_1\wedge\cdots\wedge w_p}{u_1 \wedge\cdots\wedge u_1} = \det \fnorm{w_i}{u_j}.
\end{equation}

Let us consider {\em the Hodge operator} $\star: \Ext^p(V) \to \Ext^{n-p}(V)$
\begin{equation}
 {\omega}\wedge{\star\,\vartheta } = \fnorm{\omega}{\vartheta}\, \Omega,
\quad 
 \omega, \vartheta \in \Ext(V)
\label{HodgeStar}
\end{equation}
where $\Omega \equiv e_1 \wedge \cdots \wedge e_n$ ({\em volume form}).
For $p$-form $\star \, \star = (-1)^{p(n-p)}$.
It is possible to define $d^\star : \Ext^{p}(V) \to \Ext^{p-1}(V)$ 
\begin{equation}
 d^\star = (-1)^{n(p+1)+1} \star d \star,
\label{dstar}
\end{equation}
and {\em the Laplace--Beltrami operator}
\begin{equation}
 \Delta = d^\star d + d\, d^\star.
\label{LapBel}
\end{equation}

\subsection{Algebraic Approach}

\TODO{Explain working with ``flat'' space}

Representation of the Clifford and exterior calculus may be also described in more 
algebraic way \cite{ClDir,Chev}. Let us return to consideration of an exterior 
(Grassmann) algebra, as a linear span of all possible products with 
$n$ generators $\theta_i$, if the wedge product is defined by formal equations like 
\eq{acom}
\begin{equation}
\theta_k \wedge \theta_j + \theta_j \wedge \theta_k = 0,
\quad 
\omega \wedge \vartheta = (-1)^{rs}\vartheta \wedge \omega
\quad (\omega \in \Ext^r\!,\, \vartheta \in \Ext^s),
\label{acomm}
\end{equation}
where $\Ext^r$ is the wedge product or $r$ generators.

Let us consider $n$ formal operators $\delta_i : \Ext^k \to \Ext^{k+1}$
\begin{equation}
 \delta_i : \theta_{j_1} \wedge \cdots \wedge \theta_{j_k}
 \mapsto \theta_i \wedge \theta_{j_1} \wedge \cdots \wedge \theta_{j_k}.
\label{di}
\end{equation}
If we have some matrix $g_{ij}$, it is possible to define 
the dual operators $\delta^\star_i : \Ext^k \to \Ext^{k-1}$
\begin{equation}
 \delta^\star_i : \theta_{j_1} \wedge \cdots \wedge \theta_{j_k} \mapsto
 \sum_{p=1}^k (-1)^{p-1} g_{i j_p}\,
 \theta_{j_1}\wedge\cdots\wedge\hat\theta_{j_p}\wedge\cdots\wedge \theta_{j_k},
\label{codi}
\end{equation}
where ${\hat\theta_{j_p}}$ means, that the term $\theta_{j_p}$ must be omitted.

Both $\delta_i$ and $\delta^\star_i$ belong to a space $\Lin(\Ext)$ of linear 
operators on $\Ext$. Let us consider $n$ operators
\begin{equation}
 \check\gamma_i = \delta_i - \delta^\star_i, \quad \check\gamma_i \in \Lin(\Ext),
\label{extgam}
\end{equation}
then 
\begin{equation}
 \check\gamma_i \check\gamma_j + \check\gamma_j \check\gamma_i = -2 g_{ij} \Id
\end{equation}
and an algebra generated by the operators $\check\gamma_i$ and their products ---
is the representation of the Clifford algebra $\Cl(g)$ with the quadratic form $g_{ij}$
in the space $\Lin(\Ext)$ of operators on the Grassmann algebra. The underlying vector 
space of the Clifford and the Grassmann algebras may be identified 
\begin{equation}
 \theta_{j_1} \wedge \cdots \wedge \theta_{j_k} \longleftrightarrow
 \check\gamma_{j_1} \cdots \check\gamma_{j_k},
\quad j_1 < \cdots < j_k  \le n.
\label{ClEqEx}
\end{equation}

It is possible also to write the formal Dirac operator \cite{ClDir} as some 
differential operator with coefficients in an algebra
\begin{equation}
 d = \sum_{i=1}^n \delta_i \pd{x_i},\quad
 d^\star = \sum_{i=1}^n \delta_i^\star \pd{x_i},\quad
 \Dir = d - d^\star = \sum_{i=1}^n \check\gamma_i \pd{x_i}.
\label{HodgeDir}
\end{equation}

\TODO{here really has used exterior algebra of a dual space?}

\section{Transformation Properties of Dirac Equation}
\label{Sec:DirTr}
\subsection{Spinor Representation of Lorentz Group}
\label{Sec:SpinLor}

For the Lorentz signature $(3,1)$ two Clifford algebras $\Cl(3,1)$ and $\Cl(1,3)$
are not isomorphic, but both are subalgebras of the Dirac algebra $\Cl(4,\C)$.
The algebra $\Cl(3,1)$ isomorphic with algebra of all $4 \times 4$
real matrices, it may be convenient for description of Majorana spinors,
and corresponds to possibility rewrite Dirac equation using
only matrices with real coefficients \cite{DauIV}. \TODO{... and so?}

The spinor representation $Spin(3,1)$ of the Lorentz group is isomorphic with 
a classical Lie group $SL(2,\C)$ of $2 \times 2$ complex matrices with unit 
determinant. Detailed and understanding discussion about this representation
may be found elsewhere \cite{SpinI}, but here it should be mentioned,
that such a reduction from 4D to 2D complex space is an example of ``economic'' 
representation with Spoin group discussed in \Sec{SpinPin}. 

An even subalgbera of $\Cl(3,1)$ is $\Cl(3,0) \cong M(2,\C)$, the algebra of all 
$2 \times 2$ complex matrices and it let us write usual representation of
$Spin(3,1)$ [really, $Spoin(3,1)$] using such a matrices. On the other hand,
it was noted in \Sec{SpinPin}, that not internal isomorphism \eq{SpinAct}, 
but other transformation \eq{SpoinAct} should be used in such a case.
For two-spinors it is usual expression \cite{SpinI}
$$
 \Mat[cc]{t + z & x - iy \\ x + iy & t - z} \mapsto 
 S \Mat[cc]{t + z & x - iy \\ x + iy & t - z} S^*
$$
or
\begin{equation}
 t + x \sigma_x + y \sigma_y + z \sigma_z \mapsto
 S\: (t + x \sigma_x + y \sigma_y + z \sigma_z) \: S^*
\label{PaulTransf}
\end{equation}
with $S \in SL(2,\C)$ and the conjugated matrix $S^*$.
Here a vector with four real coefficients $v=(t,x,y,z)$ maps to some 
$v'=(t',x',y',z')=L v$, with $L$ is the Lorentz transformation.  

On the other hand, for Dirac matrices and 4D complex space of Dirac spinors,
it is possible to use expression like \eq{SpinAct} with an automorphism of 
the algebra of $4 \times 4$ matrices
\begin{equation}
 t\gamma^0 + x \gamma^1 + y \gamma^2 + z \gamma^3 \mapsto
 \LA \: (t\gamma^0 + x \gamma^1 + y \gamma^2 + z \gamma^3) \: \LA^{-1},
\quad \psi \mapsto \LA \psi,
\label{GamTransf}
\end{equation}
where $\gamma^k$ are four Dirac gamma matrices \eq{GamSpin}.

Here the matrix $\LA$ is a $4 \times 4$ complex matrix, {\em e.g.} 
\begin{equation}
 \LA = \Mat[cc]{S & 0 \\ 0 & {S^*}^{-1}}, \quad S  \in SL(2,\C) 
\end{equation}
 
It was already mentioned, that the four-component Dirac spinor are useful
for description of group $O(3,1)$ with improper transformations like a
time reflection ($Pin(3,1)$ group), but transformation \eq{GamTransf} also 
convenient for covariant description of the Dirac spinors. Let us consider the 
spinor space as some abstract complex space $\C^4$, together with space of 
linear operators $\Lin(\C^4)$. 
For transition $B$ to a new basis in the space $\C^4$, may be written an
usual formula of linear algebra for transformation of operators 
\begin{equation}
 \psi' = B\psi, \quad \psi \in \C^4;
\qquad
 M' = B M B^{-1}, \quad M \in \Lin(\C^4).
\label{TransfOp}
\end{equation}
 
So, covariant properties of the Dirac operator and four-component spinors 
are in agreement with the transformations of an operator algebra.
It may be also checked directly, the transformation \eq{GamTransf} 
respects the Dirac equation 
\begin{equation}
 (\LA\Dir\LA^{-1})\LA\psi = m \LA\psi
\label{LDirL}
\end{equation}
and $\Dir \mapsto \LA\Dir\LA^{-1}$ acts on tangent vector 
({\em i.e.} partial derivatives $\partial_\mu$ in Dirac operator $\Dir$)
as a Lorentz transformation due to \eq{GamTransf}.

It should be mentioned, that \eq{TransfOp} also describes a formal transition
between different matrix representations of the generators $\gamma^k$, but it is rather 
a question of notation and does not have direct relation with the covariance. For
example, together with presentation \eq{GamSpin} may be also used
so-called {\em standard} one \cite{QFT,DauIV}  
\begin{equation}
 \gamma_s^0 = \Mat{1 & 0 \\ 0 & -1}, \quad
 \gamma_s^k = \Mat[cc]{0 & \sigma_k \\ -\sigma_k & 0} \quad (k = 1,2,3)
\label{GamStrd}
\end{equation}

Similarly with general case \Sec{Clif}, for any isometry, {\em i.e.} Lorentz
trasformation $L \in O(3,1)$ there are two ways \eq{Ae} and \eq{hah}
to write the same transformation
\begin{equation}
 \gamma^k \mapsto 
 {\gamma'}^k = \nsum_j L^k_j\gamma^j = \LA_L \gamma^k \LA_L^{-1}.
\label{LALor}
\end{equation}

It was already discussed in \Sec{Clif}, for general linear transformations of 
a basis, {\em i.e.} $G \in GL(4,\R)$ it is not possible to write analogue of 
\eq{LALor}. For description of spinors in General Relativity often is used
construction already mentioned in \Sec{ClifTrans}. It is introduced a 
diagonal metric in normalized basis together with a matrix $A$ of transition to
initial basis and metric \cite{Cart,BirDav}. The $4 \times 4$ matrix $A$ 
may be considered as four vectors called {\em tetrad} \cite{BirDav,DauII,MTBH,Asht}.

\TODO{Newman--Penrose}

\medskip

It was already mentioned, the main theme of present \paper\ --- is extension of Dirac 
equation from 4D space of Dirac spinors to 16D linear space, there covariant action 
of $GL(4,\R)$ group may be written directly. 
\TODO{... transformation between solutions of Dirac equation ... }

For such a consideration it is convenient to use different algebraic structures
on an extended linear space and here is used following sequence of 16D linear spaces
\begin{equation}
 \Ext(\C^4) \cong \Cl(4,\C) \cong M(4,\C) \cong \C^4 \otimes \C^4.
\label{AllC16D} 
\end{equation}
It is also possible to consider a real analogue of the sequence \eq{AllC16D}
\begin{equation}
 \Ext(\R^4) \cong \Cl(3,1) \cong M(4,\R) \cong \R^4 \otimes \R^4.
\label{AllR16D} 
\end{equation}

\subsection{Dirac Equation and Grassmann Algebras}
\label{Sec:DirGrass}

It was already discussed in \Sec{Ext} that exterior forms have well defined
covariant properties with respect to any transformation of coordinate
system. For an exterior algebra on the tangent space of 4D manifold $\M$
(the spacetime) we have the decomposition 
\begin{equation}
 \Ext(\M) =  \Ext^0(\M) \oplus \Ext^1(\M) \oplus \Ext^2(\M) 
 \oplus \Ext^3(\M) \oplus \Ext^4(\M)
\end{equation}
with dimensions $\{1,4,6,4,1\}$ respectively. It corresponds to decomposition
of 16D linear space $\Ext(\M)$ on five subspaces of irreducible representations
$GL(4,\R)$ group with given dimensions. Transformation properties of $\Ext^k(\M)$
corresponds to tensors with type $(0,k)$ and was already considered in \Sec{Ext}, 
see \eq{TensTransf}.

It is also convenient to consider transformation properties of $\Ext(\M)$ with respect
to $SL(4,\R)$ subgroup of the matrices with unit determinant. The group does
not change 4-volume form $\Omega = dt \wedge dx \wedge dy \wedge dz$.
The Hodge operator $\star$ \eq{HodgeStar} interchanges $\Ext^k(\M)$ and $\Ext^{4-k}(\M)$
forms. So both $\Ext^0(\M)$ and $\Ext^4(\M)$ corresponds to a trivial, scalar representation 
of $SL(4,\R)$ group, $\Ext^1(\M)$ and $\Ext^3(\M)$ --- to 4D vector representation.

Really, because $\star : \Ext^k(\M) \to \Ext^{4-k}(\M)$ acts as duality, the 
representation $\Ext^3(\M)$ should be considered as dual to $\Ext^1(\M)$.
\TODO{$\Ext^1(V) \cong V^*$, so $\Ext^3(V) \cong V$ ...}
Due to the same principle, $\star : \Ext^2(\M) \to \Ext^2(\M)$ is duality of 6D space 
$\Ext^2(\M)$. A transformation to the dual space is equivalent to the introduction of 
some metric and the Hodge duality defines a metric on 6D space $\Ext^2(\M)$.    

This metric with signature $(3,3)$ is relevant with representation of 
$SL(4,\R)$ group as double cover of $SO(3,3)$ \cite{Geom4}, two elements 
$\pm G \in SL(4,\R)$ map to same element $S \in SO(3,3)$, {\em i.e.} the 
classical (matrix) group $SL(4,\R)$ is isomorphic with $Spin(3,3)$.

It is an analogue of so-called Klein relations for Pl\"ucker coordinates and 
well known from the theory of Lie groups, twistors \cite{Post,SpinII} {\em etc.}.
The general construction for a complex space $\C^4$ and $2{\to}1$ map $SL(4,\C)$
to $SO(6,\C)$ produces by restriction on different real subspaces isomorphisms 
like $SL(4,\R) \cong Spin(3,3)$, $SU(4) \cong Spin(6)$, $SU(2,2) \cong Spin(4,2)$
{\em etc}. 

Let us consider the space of 2-forms, with coordinates denoted as 
$p^{jk} dx_j \wedge dx_k$,
then the Hodge duality corresponds to invariance of form 
\begin{equation}
 p^{01}p^{23}+p^{02}p^{31}+p^{03}p^{12}.
\label{pluform}
\end{equation}
If to introduce 6 new coordinates
\begin{equation}
\begin{array}{lll}
q_1 = (p^{01} + p^{23})/2,&
q_2 = (p^{02} + p^{31})/2,&
q_3 = (p^{03} + p^{12})/2,\\
q_4 = (p^{01} - p^{23})/2,&
q_5 = (p^{02} - p^{31})/2,&
q_6 = (p^{03} - p^{12})/2,
\end{array}
\end{equation}
the quadratic form \eq{pluform} may be rewritten as
\begin{equation}
 q_1^2+q_2^2+q_3^2-q_4^2-q_5^2-q_6^2.
\label{plu33}
\end{equation}
New basis in the space $\Ext^2(\M)$ corresponding to such a diagonal form may be
chosen as
\begin{equation}
\begin{array}{ll}
\eta_1 = dx_0 \wedge dx_1 + dx_2 \wedge dx_3, &
\eta_4 = dx_0 \wedge dx_1 - dx_2 \wedge dx_3, \\ 
\eta_2 = dx_0 \wedge dx_2 + dx_3 \wedge dx_1, &
\eta_5 = dx_0 \wedge dx_2 - dx_3 \wedge dx_1, \\
\eta_3 = dx_0 \wedge dx_3 + dx_1 \wedge dx_2, &
\eta_6 = dx_0 \wedge dx_3 - dx_1 \wedge dx_2.
\end{array}
\label{plux2}
\end{equation}
If a basis of 1-forms $dx_k \in \Ext^1(\M)$, $k=0,\ldots,3$ is transformed by element of 
group $SL(4,\R)$, then the basis of 2-forms $\eta_p \in \Ext^2(\M)$, $p=1,\ldots,6$ is 
transformed by some element of $SO(3,3)$, because it saves \eq{plu33} invariant, 
and it is also clear from \eq{plux2}, two 
elements $\pm G \in SL(4,\R)$ correspond to the same transformation.

\medskip

So action of the group $SL(4,\R)$ on 16D space $\Ext(\M)$ is decomposed on trivial 
representation for 1D spaces $\Ext^0(\M)$ and $\Ext^4(\M)$, 4D ``vector'' representation 
$SL(4,\R)$ on spaces $\Ext^1(\M)$ and $\Ext^3(\M)$, and 6D representation $SO(3,3)$ on 
``autodual'' space $\Ext^2(\M)$.   

Full group $GL(4,\R)$ also includes the dilations $\lambda \Id$ (where 
$\lambda\in \R$, $\lambda > 0$ and $\Id$ is $4 \times 4$ unit matrix) and the
improper transformations with $\det = -1$, like time inversion. The dilations acts as 
$\lambda^k$ on $\Ext^k(\M)$. For any improper transformation there is 
additional $(-1)$ multiplier for $\Ext^4(\M)$ (``pseudoscalars''), $\Ext^3(\M)$ 
(``pseudovectors'') and quadratic form in \eq{plu33}.

It was already discussed in \Sec{Hodge}, the Dirac operator on exterior forms
\newline $\check\Dir : \Ext(\M) \to \Ext(\M)$ 
is defined in covariant way \eq{HodgeDir}
\begin{equation}
 \check\Dir  = (d - d^\star), \quad
 d : \Ext^k(\M) \to \Ext^{k+1}(\M), \quad 
 d^\star : \Ext^k(\M) \to \Ext^{k-1}(\M).
\end{equation}
Here for definition of the operator $d^\star$ it is necessary to have a metric $g$
on the vector space $V = T\M$ (tangent space of $\M$), and if there are given 
$\Upsilon \in \Ext(V)$, $g$, $d$ and $d^\star$
in one coordinate system, it is possible to write transformation of all four 
geometrical objects for an arbitrary change of the coordinate system. So if
the Dirac-K\"ahler equation 
\begin{equation}
 \check\Dir \Upsilon \equiv (d - d^\star) \Upsilon = m \Upsilon,
\tag{\ref{DirForm}$'$}
\label{DirKahl}
\end{equation}
was satisfied in one coordinate system, it is also true for any transformation.
So \eq{DirKahl} is covariant with respect to $GL(4,\R)$ group of general linear
transformations of coordinate system.

It should be mentioned, both $d$ and $d^\star$ change range of homogeneous
$k$-forms from $\omega \in \Ext^k(V)$ and so the form $\Upsilon \in \Ext(V)$ might be 
{\em nonhomogeneous} to satisfy \eq{DirKahl}. Such a form is a sum of
few $k$-forms with different $k$ and it is enough to write transformation
for each such a term.

The relation between such a ``tensor'' form of a Dirac equation and spinoral one 
is well known \cite{Marchuk,DirHes}, but the purpose of given \paper\ is to consider 
{\em transformation properties} of the equations and a model of Dirac spinor as a 
{\em subsystem} in a 16D space. It is convenient first to revisit a model of Dirac 
spinors as an ideal of Clifford algebra \cite{DirHes,QuatDir}.

\subsection{Matrix Dirac Equation and Clifford Algebras}
\label{Sec:DirClif}

Dirac equation on a Clifford algebra may be simply written
\begin{equation}
 \Dir \mPsi = m \mPsi, \quad \mPsi \in \Cl(4,\C) \cong M(4,\C), 
 \quad \Dir = \sum_{k=0}^3 i \gamma^k \pd{x_k}.
\label{mxDirEq}
\end{equation}
Here $\mPsi$ is not spinor, but element of the Clifford algebra. The Spin group also
may be considered as some subset of the Clifford algebra, so $\gamma^k$ and $\mPsi$
are elements of the same space $\Cl(4,\C)$. 

Formally usual Dirac equation \eq{DirEq} may be considered as special form of
\eq{mxDirEq}, then the matrix $\mPsi$ has only one nonzero column $\psi$. More
algebraic way to describe such a model --- is to consider {\em left ideals}
of a Clifford algebra \cite{DirHes}.

The left ideal of an algebra $\mathcal A$ by definition \cite{SLang}
is a linear subspace $\mathcal I \subset \mathcal A$ with property
$\mathcal{A I} \subset \mathcal I$, {\em i.e.} any element of the algebra after
multiplication on an element of the ideal produces again an element of the ideal.
A simplest example of a left ideal in the matrix algebra is a set of matrices
with only one nonzero column already mentioned above
\begin{equation}
 M_\psi = \Mat{\psi_1 & 0 & 0 & 0 \\ \psi_2 & 0 & 0 & 0 \\
\psi_3 & 0 & 0 & 0 \\ \psi_4 & 0 & 0 & 0}
\label{MatOneCol}
\end{equation}

A specific point of such representation is the transformation properties of 
given equation. All elements of an algebra should have the same transformation
properties, so if to consider $\mPsi$ and $\gamma^k$ as equivalent elements
of the same algebra, then even for the Lorentz transformations instead of \eq{LDirL}
with $\psi \mapsto \LA \psi$, it is necessary to consider
\begin{equation}
(\LA\Dir\LA^{-1})\LA\mPsi\LA^{-1} =  m \LA\mPsi\LA^{-1},
\quad \mPsi \mapsto \LA\mPsi\LA^{-1}.
\label{LmPsiL}
\end{equation}

Such a difference in a transformation property of $\mPsi$ does not produce a 
serious problem for comparison with the initial equation \eq{DirEq}, due to
an additional symmetry $\mPsi \to \mPsi R$ of \eq{mxDirEq}, {\em i.e.} the 
right multiplication on the arbitrary element $R$ of the Clifford algebra.    

In such a case the right multiplication on $\LA^{-1}$ in \eq{LmPsiL} is not so
essential. For the comparison of \eq{mxDirEq} with a spinor equation \eq{DirEq},
it is necessary to consider not only the element $M_\psi$ \eq{MatOneCol}
of some left ideal, 
but also all $M_\psi R$, {\em i.e.} space of matrices with all columns 
proportional to the same vector. 

If $\psi \in \C^4$ is the initial 4-spinor, and 
$\alpha = (\alpha_1,\alpha_2,\alpha_3,\alpha_4) \in \C^4$ is
an arbitrary vector of coefficients, then instead of \eq{MatOneCol}
in is necessary to consider a matrix with proportional columns, 
$M_{ij}=\psi_i\alpha_j$,
\begin{equation}
 M = \Mat{\alpha_1\psi_1 & \alpha_2\psi_1 & \alpha_3\psi_1 & \alpha_4\psi_1\\
\alpha_1\psi_2 & \alpha_2\psi_2 & \alpha_3\psi_2 & \alpha_4\psi_2\\
\alpha_1\psi_3 & \alpha_2\psi_3 & \alpha_3\psi_3 & \alpha_4\psi_3\\
\alpha_1\psi_4 & \alpha_2\psi_4 & \alpha_3\psi_4 & \alpha_4\psi_4}, 
 \quad M = \psi \alpha^{T} \equiv \psi \otimes \alpha.
\label{bipsi}
\end{equation}
For arbitrary matrices $L,R$ it may be written
\begin{equation}
  LMR = L\,(\psi \otimes \alpha)\,R = (L\psi) \otimes (R^{T} \alpha),
\label{LMR}
\end{equation}
so the left and right multiplication saves ``the product structure'' and it
will be used further in \Sec{DirTens}, but space of all matrices in form
\eq{bipsi} is not an ideal, because it is not {\em linear subspace}, {\em e.g.}
sum of elements does not necessary may be presented again as a tensor product
of two vectors \eq{bipsi}. Really linear span of the ``degenerate'' ($\det M = 0$) 
subspace of all such matrices \eq{bipsi} coincides with the whole matrix algebra.

Yet, fixed $\alpha \in \C^4$ corresponds to {\em the left ideal}, for example initial 
construction with one column is the particular case of such an ideal for 
$\alpha = (1,0,0,0)$. On the other hand, fixed $\psi$ corresponds to a linear 
subspace, {\em the right ideal} $\mathcal R_{\psi}$ of the algebra. 

\smallskip

Such a procedure also has analogues with consideration of {\em an equivalence classes} 
$M R \sim M$, $M \in \Cl$ \cite{DirHes} and partially with usual spinor case, where 
different physical states correspond to {\em the rays in Hilbert space}, {\em i.e.} 
equivalence classes $\psi \sim \lambda \psi$. 
A model for relation between Dirac spinors $\psi \in \C^4$ and the equivalence 
classes in 16D linear spaces $\Cl(4,\C) \cong M(4,\C) \cong \Ext(\C^4)$ is called 
{\em Dirac--Hestenes spinors} in \cite{DirHes}. On the other hand, only $SO(3,1)$ 
group respects the equivalence classes and so such construction is not valid for 
general coordinate transformations from $GL(4,\R)$ \cite{GrDir03}. 

The property should be discussed here with more details. The covariant property of
$\Ext(\R^4)$ or $\Ext(\C^4)$ discussed in \Sec{DirGrass} may be extended to other
linear spaces $\Cl(4,\C) \cong M(4,\C)$ if to choose an isomorphism of the linear
spaces (without algebraic structure) and to use a diagram
\begin{equation}
\begin{array}{ccc}
 \Dir\mPsi &\longrightarrow&  \Dir'\mPsi' \\
\big\downarrow{\rlap{$^\ClEx$}}&  & \big\uparrow{\rlap{$^{\ClEx^{-1}}$}} \\
 \check\Dir\Upsilon &\longrightarrow& \check\Dir'\Upsilon' 
\end{array}
\label{DiaClEx}
\end{equation}

Here the map of the linear spaces $\ClEx : \Cl \to \Ext$ for the diagonalized 
(Lorentz) metric is defined as \eq{ClEqEx}
\begin{equation}
 \ClEx : \gamma^{i_1}\cdots\gamma^{i_k} \mapsto 
 dx_{i_1} \wedge \cdots \wedge dx_{i_k}
\label{dClEx}
\end{equation}
It is convenient also to choose the particular matrix representation of Clifford 
algebra say \eq{GamSpin}, and to consider \eq{mxDirEq} as a matrix equation,
and let $\ClExm$ is the map between the linear spaces of $4 \times 4$ 
matrices and exterior forms, corresponding to $\ClEx$ in the given matrix 
representation. Then the transformation of the matrix $\mPsi$ 
for arbitrary $G \in GL(4,\R)$ may be written
\begin{equation}
 G_M : \mPsi \mapsto \ClExm^{-1}\Bigl(G_\Ext\bigl(\ClExm(\mPsi)\bigr)\Bigr),
\label{GMat}
\end{equation}
where $G_\Ext$ is the action $G$ on the space $\Ext$ of the nonhomogeneous forms 
discussed above in \Sec{DirGrass}. 

The \eq{GMat} sometime is better to consider as a formal equation for a map
between two linear spaces without any additional structures, {\em i.e.}
the choice of gamma matrices and bases in $M(4,\C)$ and $\Ext(\C^4)$ represent
the map $\ClExm$ and $G_\Ext$ as some formal linear transformations of $\C^{16}$.
The map $G_M$ may be considered as the composition of the transformations 
$G_M = \ClExm \, G_\Ext \, \ClExm^{-1}$, {\em viz} product of three $16 \times 16$ 
complex matrices.

Such a ``technical'' point of view is justified for representation of the general 
coordinate transformations, because $G_M$ is the isomorphism only for the Lorentz 
group, but for a general element $G \in GL(4,\R)$, it is not necessary in agreement 
with Clifford multiplication
\begin{equation}
 \exists\, G \notin SO(3,1) : \quad 
 G_M(\gamma^j)G_M(\gamma^k) \ne G_M(\gamma^j\gamma^k), 
\label{GMneq}
\end{equation}
and so attempts to use a definition like \eq{dClEx} for a Clifford algebra with 
a non-diagonal quadratic form $g$ may produce some problems. Formally \eq{dClEx}
should be chosen for a normalized basis. \TODO{... that is for non-normal bases?} 

The inequality \eq{GMneq} appears because only isometries always respect the subspaces 
of $k$-multivectors $\Cl^{(k)}$ \eq{Clk}, but the subspaces of $k$-forms $\Ext^k$
due to \eq{TensTransf} are invariant for arbitrary linear transformations.
So $G_M(\gamma^{i_1}\cdots\gamma^{i_k}) \in \Cl^{(k)}$, but 
$G_M(\gamma^{i_1})\cdots G_M(\gamma^{i_k})$ may include also additional terms 
from $\Cl^{(j)}$, $j \le k-2$.

\medskip
\noindent {\em It is useful to consider some examples.} 
\begin{itemize}
\item 
Let we have some transformation in $x-y$ plane.
\begin{equation}
 \gamma^1 \mapsto a_{11} \gamma^1 + a_{12} \gamma^2, \quad
 \gamma^2 \mapsto a_{21} \gamma^1 + a_{22} \gamma^2,
\end{equation}
then the multivector $\gamma^1\gamma^2$ is transformed as
$$
 \gamma^1\gamma^2 \mapsto 
  -(a_{11}a_{21}+a_{12}a_{22})\Id + (a_{11}a_{22}-a_{12}a_{21}) \gamma^1\gamma^2
$$
So a rotation $a_{11}=a_{22} = \cos{\varphi}$, $a_{12} = - a_{21} = \sin{\varphi}$
maps $\gamma^1\gamma^2$ to itself, but for a nonorthogonal transformation an
additional term with unit appears. On the other hand, for the 2-form $dx_1 \wedge dx_2$
such a term may not appear, because $dx_1 \wedge dx_1 = dx_2 \wedge dx_2 = 0$.
Only for rotations $\gamma^1\gamma^2$ and $dx_1 \wedge dx_2 = \ClEx(\gamma^1\gamma^2)$ 
have the same transformation properties.

\item
On the other hand, for a transformation in $t-x$ plane
\begin{equation}
 \gamma^0 \mapsto b_{00} \gamma^0 + b_{01} \gamma^1, \quad
 \gamma^1 \mapsto b_{10} \gamma^0 + b_{11} \gamma^1,
\label{txlin}
\end{equation}
then the multivector $\gamma^0\gamma^1$ is transformed as
$$
 \gamma^0\gamma^1 \mapsto 
  (b_{00}b_{10}-b_{01}b_{11})\Id + (b_{00}b_{11}-b_{10}b_{01}) \gamma^0\gamma^1
$$
So, the boosts $b_{00}=b_{11}=\cosh{\upsilon}$, $b_{10}=b_{01}=\sinh{\upsilon}$ map
$\gamma^0\gamma^1$ to itself, {\em i.e.} again only the isometries of the concrete 
quadratic form (the pseudo-Euclidean Minkowski metric) respect the subspaces
of the Clifford $k$-multivectors ($\Cl^{(2)}$ in given examples).   
\end{itemize}

\TODO{... to prove: invariance of $\Cl^{(k)}$ if (and only if ?) isometry}

\subsection{Tensor Product of Two Spinor Spaces}
\label{Sec:DirTens}
 
For any element of Lorentz group $L \in O(3,1)$ exists spinor representation 
\eq{LmPsiL} of transformation \eq{GMat}, {\em i.e.}
\begin{equation}
 \forall L \in O(3,1), ~  \exists \LA_L: \quad 
 L_M (\mPsi) = \LA_L \mPsi \LA_L^{-1}.
\label{yaLmL}
\end{equation}
On the other hand, any linear transformation of the 16D space of $4 \times 4$ 
matrices may be represented as 
\begin{equation}
 \mPsi \to \sum_J A_J \mPsi B_J,
\label{sumAB}   
\end{equation}
there $A_J$, $B_J$ are some matrices. So for arbitrary $G \in GL(4,\R)$ 
the transformation $G_M$ also may be represented using \eq{sumAB}, but only
for Lorentz transformation such representation has only one term \eq{yaLmL}.

It has an analogue with idea of representation $\Cl$ as tensor product
of two spinor spaces $S \otimes S$ in general algebraic theory \cite{Chev}.
For present discussion it is enough to consider concrete model with
algebra of $4 \times 4$ complex matrices. It corresponds to last term 
\begin{equation}
M(4,\C)\cong\C^4 \otimes \C^4
\label{SxS} 
\end{equation}
in sequence of 16D linear spaces \eq{AllC16D} already mentioned briefly 
\Sec{DirClif} in relation with treatment of solution of the matrix Dirac 
equation \eq{mxDirEq} as a tensor product of two complex 4-vectors \eq{bipsi}. 

Due to \eq{yaLmL}, for a transformation $L$ from Lorentz group \eq{LMR} 
may be rewritten as  
\begin{equation}
 L_M(\psi\otimes\alpha) = (\LA_L \psi) \otimes ({\LA_L^{-1}}^T \alpha),
\quad \psi,\alpha \in \C^4,
\label{LocLor}
\end{equation}
but for a general linear transformation $G \in GL(4,\R)$ it is necessary
to use \eq{sumAB} with some set $\mathfrak{L}_J(G)$, $\mathfrak{R}_J(G)$
\begin{equation}
 G_M(\psi\otimes\alpha) = \sum_J(\mathfrak{L}_J \psi) \otimes (\mathfrak{R}_J \alpha),
\quad \psi,\alpha \in \C^4,
\label{EntGL}
\end{equation}

\medskip

Technically the \eq{EntGL} even not guarantee the existence of an unique 
decomposition and may look more complicated, than the initial equation
\eq{GMat}. Method of calculation \eq{GMat} is difficult, but straightforward,
it would be possible to write precise expressions for any choice of isomorphism
of 16D linear spaces $\Ext(\C^4)$ and $M(4,\C) \cong \C^4 \otimes \C^4$ ({\em c.f.}
\cite{Marchuk,DirHes}) using tedious calculations or cumbersome output of some 
computer algebra system, but it most likely would not clarify a physical properties 
of the transformation. On the other hand, the \eq{EntGL} may be useful for some 
intuitive explanation \cite{GrDir03} used in further discussion. 

\subsection{A Formal Model with Two Quantum Systems}

In the quantum mechanics the state space of a compound quantum system may be 
described as the tensor product of the spaces of each subsystem. So, formally due 
to \eq{SxS} $\mPsi$ may be considered as a composite system with two particles. 
The parts are called {\em not entangled} (uncorrelated), if the state such a 
system may be represented as a tensor product $\psi_1 \otimes \psi_2$ \cite{Peres}. 
So, in \eq{bipsi} $\mPsi$ is a formal analogue of such non-entangled system 
$\psi \otimes \alpha$. It is clear also, that due to \eq{LocLor} a Lorentz 
transformations respect the entanglement. {\em The correspondence between 
$\mPsi$ and $\psi$ is an analogue of relation between system and subsystem in
the quantum mechanics.}

It should be mentioned, that in the constructions used below $\psi(x)$ is 
a function of the point $x$, but $\alpha$ is a constant complex 4-vector.  
Formally $\mPsi$ corresponds to the two uncorrelated systems, $\psi$ and $\alpha$. 
The first system satisfies to Dirac equation. The second one formally also satisfies 
Dirac equation for a massless particle, but a state of the system does not matter:
it is not correlated with the first one, can be considered as an auxiliary system 
and a Lorentz transformation may not change such a structure. So, if to consider
only the Lorentz transformations, the matrix Dirac equation \eq{mxDirEq} is 
equivalent with usual one \eq{DirEq}. 

On the other hand, a general coordinate transformation due to \eq{EntGL} does not
preserve the property of being non-entangled. Formally we may reduce the matrix Dirac 
equation \eq{mxDirEq} to usual one \eq{DirEq} by restriction to an equation for
the arbitrary row of $\mPsi$, but after the general coordinate transformation each 
column of $\mPsi$ contains combinations of $\psi_i$ with different $\alpha_j$, 
and so depends on the state of the second system. It differs from the case with 
Lorentz transformations, then any combination of $\alpha_j$ for any particular 
column due to \eq{LocLor} appears as a common scalar multiplier and always may
be omitted.

In the quantum mechanics there is a standard method of consideration of a subsystem
of an entangled system --- the mixed states and the density 
matrix \cite{Peres,DauIII,DauV}. It should be recalled, that all the structures 
discussed here in relation with the formal decomposition $\mPsi \in S \otimes S$ 
yet have {\em rather mathematical} resemblance with the quantum mechanics and was 
used mainly with an illustrative reason, but the appearance of some analogue of 
the mixed states directly from the properties of the group $GL(4,\R)$
of the general linear transformation of coordinate systems, may be promising.  

\section{Curved Space-Time and Singularities}
\label{Sec:GR}

\subsection{Dirac Operator and General Relativity}
\label{Sec:DirGR}

Despite of using an arbitrary metric and the general linear transformation of 
a coordinate system, the vast amount of the material presented above was 
relevant rather with the flat space-time. The applications of the Dirac operator 
in a curved space may depend on some implicit suggestions. For example 
for the Euclidean metric it is possible to calculate different kinds of 
local Laplacians $\Dir^2$: spinor, Hodge or twisted one \cite{ClDir}. 
All such Laplacians contain a term with scalar curvature $\frac{1}{4}R(x)$. 
The Hodge and twisted spinor Laplacians also contain additional terms with 
sums of fourth order \cite{ClDir}. Such a kind of terms could produce a specific 
kind of the Klein--Gordon operator in the curved space time like
$\KG \pm \frac{1}{4}R + m^2$,
and it does not look as a proper physical one.

It was already mentioned that most calculations with the spinors in General Relativity
use the various applications of tetrad formalism or some analogues. 
For example the Dirac equation in the Newman--Penrose formalism for a charged 
black hole described by the Reissner--Nordst\"orm metric and for a rotated one 
with the Kerr metric may be found in \cite{MTBH}. 

On the other hand, there is some formal question:
{\em it is known, that the black hole evaporation may be related with a 
transition from a pure state to a mixed state of a quantum system 
{\rm \cite{BirDav,Isham,HawkPen,Beken}}, but how in principle to consider 
such a transition?}. 

It was mentioned at end of \Sec{DirTens}, that analogue of such transition 
has some allusions with consideration of the general linear coordinate 
transformation on the extended (matrix) Dirac equation \eq{mxDirEq}.
 
Of course, there are huge amount of works and different approaches to black
hole evaporation. It is necessary to use methods of field theory for rigour
research \cite{BirDav,PerTer,Wern,Terno}, but such methods in very core always
use covariance with respect to group of coordinate transformations, an so it 
is anyway necessary to consider structure of such a group first, for
using theory of infinite-dimensional unitary representations \cite{AQFT,ThGr}.
Theory of infinite-dimensional representation of $GL(4,\R)$ and affine groups
in relation with generalization of Dirac equation is also discussed in \cite{Kir}.

It should be mentioned also, that here is considered property of single system.
In some works like \cite{PerTer} transition from pure to mixed
state in black hole, ``information loss paradox'' is considered as an analogue 
of process with ``intervention of classical world'' in modern applied
theory of open quantum systems\footnote{Convenient for practical applications,
but partially empirical, with lack of agreement on some fundamental theoretical
questions.}, but in such a case the process of evaporation of black holes lost 
the uniqueness attracting much attention. Introduction of density matrix, {\em i.e.} 
statistical operator in very beginning suggests incomplete description of quantum 
systems \cite{DauII,DauV} and so in such a case instead of an answer the question 
partially transferred in set of postulates. 

On the other hand, the considered model of the matrix Dirac equation justifies the
phenomenon of transition from pure to mixed state not for wide variety of conditions, 
but only in presence of a singularity (see \Sec{GaugeGR} below), and so may provide 
the subtler classification. Despite of the technical differences, here the treation of
mixed state resembles \cite{DeuTM}, {\em i.e.} it {\em does not suggest} with very 
beginning neither the work with statistical ensembles, nor interaction with 
environment, nor any ideas about relation between quantum and classical world, 
{\em etc}. The lost of information in present model is due to consideration of
only four components $\psi$ in $4 \times 4$ matrix $\mPsi$.

\subsection{Connections, Gauge Theories and Gravitation}
\label{Sec:GaugeGR}

The problem of the consideration of spaces with singularities is well 
known \cite{HawkEl}. If such a problem may have some relevance with a
transition from a pure to a mixed state due to usage of an extended covariant 
equation \eq{mxDirEq} and $GL(4,\R)$ group discussed above? 

For a space with a singularity there is a problem with definition of metric on some 
subspaces.
The (pseudo-)Riemannian manifold is the particular example of the affinely connected 
space \cite{KobNomI}. The affine or linear connection and the metric --- are two 
different geometrical objects, it is possible formally to consider a manifold 
with linear connection without any metric at all\footnote{It is should not be 
treated as a ``flat'' space, because the conception of flatness or curvature 
is not possible to introduce without a metric.}. 

The affinely connected space also is the particular example of a more general 
theory of connections on a principle bundle with an arbitrary structure group, 
but this theory has some counterpart in the physics --- {\em the gauge theory}. 
For example such a link is especially transparent in {\em Ashtekhar's (loop) 
quantum gravity} \cite{Asht}, {\em Ivanenko's gauge theory of gravity} \cite{IvPS}, 
but many methods may be useful for description of different structures in 
arbitrary models of gravitation, including a ``canonical'' one, higher-dimensional 
and super-gravitation \cite{SardI}.

\smallskip
 
In the general theory, it is considered a connection on {\em a principal bundle} 
with some Lie group \cite{KobNomI}. For a linear connection the structure group 
is $GL(4,\R)$, for affine one --- the affine group $A(4)$, but it is possible to 
work with the linear connection on a manifold instead of affine one without 
lost of generality \cite{KobNomI}. The important question --- is the reduction of 
the structure group to some subgroup. 

For General Relativity, it is the reduction to the Lorentz group $SO(3,1)$ --- 
it is just the question, if it is possible to choose {\em the atlas}, there all 
transition between different maps may be described by transformations from the 
considered subgroup \cite{KobNomI}. In General Relativity it is related with 
the question about possibility to use only the transformation from the Lorentz group 
for transition between different coordinate systems and due to a general 
theorem of differential geometry \cite{KobNomI},
it is possible iff {\em globally} exist the Minkowski metric 
and the tetrad field --- it is one 
geometrical treatment of {\em the equivalence principle in General Relativity}
\cite{IvPS,SardI}.

The consideration above shows, {\em that the questions about a singularity of
the metric and about impossibility of the reduction from the general coordinate 
transformation to Lorentz one are directly related.} Of course there is a 
problem, if the physical singularity does not accept also introduction of 
an arbitrary group of transformation, even more general than Lorentz group. 
A ``toy model'' discussed below in \Sec{Schw} may be useful from such 
a point of view.

\subsection{Schwarzschild Spacetime}
\label{Sec:Schw}

Initial metric of Schwarzschild spacetime may be represented in polar 
coordinates as \cite{DauII,MTBH,HawkEl}
\begin{equation}
 ds^2 = (1 - r_g/r) dt^2 - (1 - r_g/r)^{-1}dr^2 - r^2 dS^2,
\quad r > r_g
\label{Schw}
\end{equation}
where $dS^2 = d\theta^2 + d\varphi^2 \sin^2(\theta)$ is the usual spherical 
surface metric and $r_g$ is a constant (the gravitational radius). If the 
metric used for the description of the gravitation field outside of a 
spherical body, it should be chosen $r_g = 2\kappa m/c^2$ to satisfy 
a Newtonian limit \cite{DauII}. Let us ``rescale'' $r_g = 1$ for convenience.
\begin{equation}
 ds^2 = \frac{r - 1}{r} dt^2 - \frac{r}{r - 1}dr^2 - r^2 dS^2,
\quad r > 1.
\tag{\ref{Schw}$'$}
\label{Schw1}
\end{equation}

\begin{figure}[ht]
\hrule
\smallskip
\begin{center}
\includegraphics{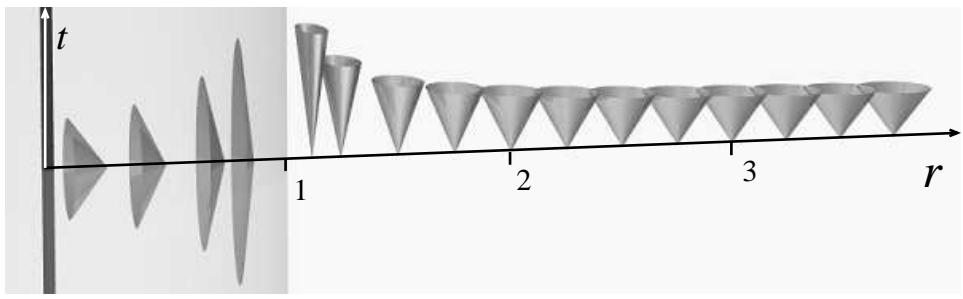}
\end{center}
\caption{Light cones for Schwarzschild metric.}\label{Fig:metSchw}
\hrule
\end{figure}

If try to extend the metric \eq{Schw} for arbitrary values $r \ge 0$,
it has singularities at $r=0$ and $r=1$. The usual practice is to consider 
a question about the isometric embedding of the space $r>1$ as a some 
subspace of a bigger manifold \cite{HawkEl}.

Here is useful to recall a wide family of such embeddings described as \cite{DauII}
\begin{equation}
 \tau = \pm t \pm \int \frac{r f(r)}{r-1}dr, \quad
 R = t + \int \frac{r}{(r-1)f(r)}dr,
\label{DauFink} 
\end{equation}
there $f(r)$ is some function of $r$. For example the Eddington--Finkelstein 
coordinates is described \cite{HawkEl,FinkCoor} 
\begin{equation}
 v = t + \tilde r, \quad w = r - \tilde r,
\quad \tilde r \equiv \int \frac{r}{r-1}dr = r + \ln(r-1),
\end{equation}
formally corresponding to the choice $f(r)\equiv 1$. For such a case in the
coordinates $(v,r,\theta,\varphi)$ the metric may be written
\begin{equation}
ds^2 = \frac{r - 1}{r} dv^2 - 2 dr dv - r^2 dS^2.
\label{EddFink}
\end{equation}
In \cite{DauII} is chosen $f(r) = 1/\sqrt{r}$, because in such a case the
coordinate system $(\tau,R,\theta,\varphi)$ is synchronous ($g_{\tau\tau}=1$).

It is convenient here to use yet another choice: $f(r) = \sqrt{2r-1}/r$,
then \eq{DauFink} produce 
\begin{eqnarray}\label{CooRot}
  \tau &=&  t - \int{\!\frac{\sqrt{2r-1}}{r-1}\,dr} 
 = t - 2 \sqrt{2r-1} + \ln\Bigl(\frac{\sqrt{2r-1}+1}{\sqrt{2r-1}-1}\Bigr), \\
 R &=& t + \int{\!\frac{r^2 \: dr}{(r-1)\sqrt{2r-1}}} =
 t + \frac{r+4}{3}\sqrt{2r-1} 
 -\ln\Bigl(\frac{\sqrt{2r-1}+1}{\sqrt{2r-1}-1}\Bigr).
 \nonumber
\end{eqnarray}
It is possible similar with \eq{EddFink} to write a metric \eq{Schw1} in 
the ``mixed'' (``skew'') coordinates $(\tau,r,\theta,\varphi)$, 
$dt = d\tau + \sqrt{2r-1}/(r-1)\:dr$
\begin{equation}
ds^2 = \frac{r - 1}{r} (d\tau^2 - dr^2) + \frac{2 \sqrt{2r-1}}{r} dr d\tau - r^2 dS^2,
\quad r \ge \frac{1}{2}.
\label{ERotFink}
\end{equation}
The metric \eq{ERotFink} may be rewritten as
\begin{equation}
(\cos \vartheta(r) d\tau + \sin \vartheta(r) dr)^2 - 
(\cos \vartheta(r) dr - \sin \vartheta(r) d\tau)^2 - r^2 dS^2,
\label{CSRotBH}
\end{equation}
where
\begin{equation}
\vartheta(r) = \mathrm{arccosec}(\sqrt{2r}),
\qquad \sin \vartheta(r) = \frac{1}{\sqrt{2r}},
\quad \cos \vartheta(r) = \frac{\sqrt{2r-1}}{\sqrt{2r}}.
\label{AngRotBH}
\end{equation}
 
\begin{figure}[ht]
\hrule
\smallskip
\begin{center}
\includegraphics{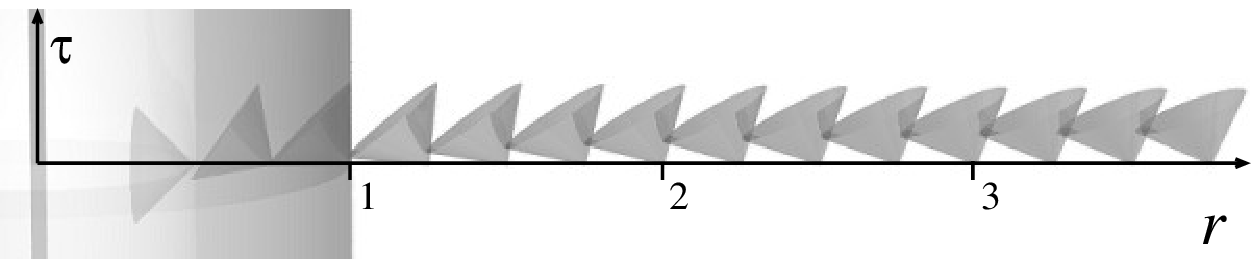}
\end{center}
\caption{Light cones in \eq{CSRotBH} metric.}\label{Fig:metrot}
\hrule
\end{figure}

The transformation of a local frame
\begin{equation}
 d\tau  \mapsto (\cos \vartheta d\tau + \sin \vartheta dr),
\quad
 dr \mapsto (\cos \vartheta dr - \sin \vartheta d\tau),
\label{LocERot} 
\end{equation}
used in \eq{CSRotBH} just corresponds to utilization of the
linear transformations more general, than Lorentz one. In \eq{txlin} \Sec{DirClif}
below already was considered an example of a general linear transformations in $t-x$ 
plane. The \eq{CSRotBH} is similar with the case, then instead of the isometric 
transformations in tangent space, {\em i.e.} boost from $SO(3,1) \subset GL(4,\R)$ 
with $b_{00}=b_{11}=\cosh{\upsilon}$, $b_{10}=b_{01}=\sinh{\upsilon}$, it is 
considered an element of $SO(4) \subset GL(4,\R)$ with $b_{00}=b_{11}=\cos{\vartheta}$,
$b_{10}=-b_{01}=\sin{\vartheta}$. It is always exist reduction of $GL(4,\R)$ to maximal
compact subgroup $SO(4)$ \cite{KobNomI,SardI}. 

It is important to mention, that despite of possibility to introduce
a local Lorentz metric for any point, it does not possible to do it globally,
because it must change sign at some moment ($R$ ``becomes time'' instead
of $\tau$) and it is responsible for appearance of horizon and discontinuity
for metric \eq{ERotFink} in diagonal form\footnote{{\em C.f.} expression for 
general $f(r)$ in \cite{DauII}.} with $(\tau,R,\theta,\varphi)$
\begin{equation}
 ds^2 = \frac{r(\tau,R)}{r(\tau,R)-1}\left(d\tau^2 
  - \frac{2r(\tau,R)-1}{r^2(\tau,R)}dR^2\right) - r^2(\tau,R)\, dS^2,
\quad r(\tau,R) \ge \frac{1}{2}. 
\label{ERotR}
\end{equation}

The coordinates \eq{CooRot} was chosen to write a simpler 
example of general linear coordinate transformation. It could be possible to
choose some other function $f(r)$, to prevent limitation $r \ge 0.5$, but really
presented metric may be defined globally. It is enough to merge two asimptotically
flat spaces 
\begin{equation}
ds^2 = \frac{r - 1}{r} (d\tau^2 - dr^2) \pm \frac{2 \sqrt{2r-1}}{r} dr d\tau - r^2 dS^2,
\quad r \ge \frac{1}{2}
\label{ERotWH}
\end{equation}
and produce an analogue of {\em wormhole} \cite{WHWEC,BHTW,KrasnWH}. 
Formally in such a case we let wider diapason of ``Euclidean rotation'' \eq{LocERot}
$0 \le \vartheta \le \pi$ instead of $0 \le \vartheta \le \pi/2$ for in 
initial metric \eq{ERotFink}. 

\begin{figure}[ht]
\hrule
\smallskip
\begin{center}
\includegraphics{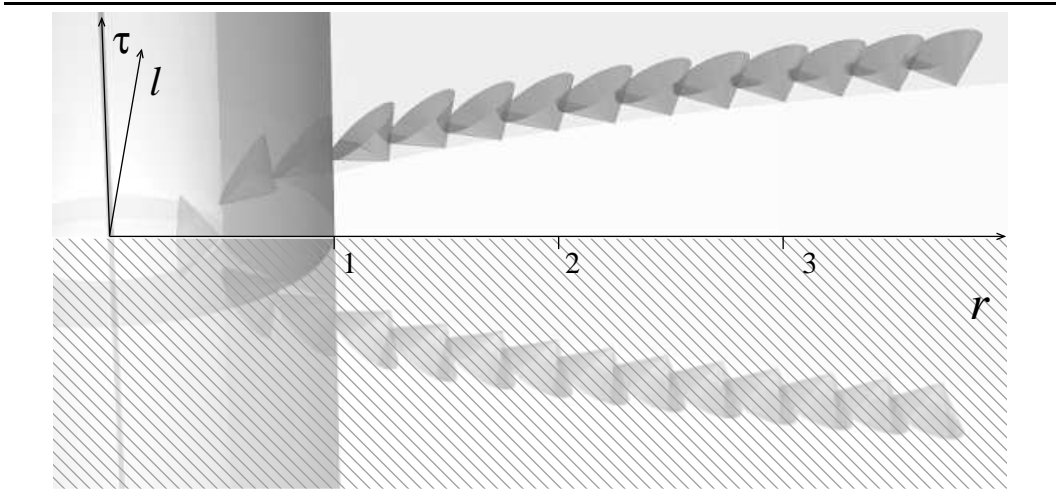}
\end{center}
\caption{Light cones for wormhole metric \eq{ERotWH}, \eq{ERotWHl}. 
A try to show five-dimensional embedding $(\tau,r,l,\theta,\varphi)$ on 
a two-dimensional picture (see also color picture on the title page).}
\label{Fig:metrotwh}
\hrule
\end{figure}

The model of wormhole \eq{ERotWH} does not look smooth for $r=1/2$, but it is  
possible to consider a formal embedding of the four-dimensional manifold 
$(\tau,r,\theta,\varphi)$ in a five dimensional one $(\tau,r,l,\theta,\varphi)$,
where $r=(l^2+1)/2$, $l = \pm \sqrt{2r-1}$ and show, that it is simply
due to a problem with a choice of the coordinate $r$ ``perpendicular to the 
throat of the wormhole.'' Using coordinates $(\tau,l,\theta,\varphi)$, it is 
possible to rewrite \eq{ERotWH}
\begin{equation}
 ds^2 = \frac{l^2-1}{l^2+1}(d\tau^2-l^2 dl^2) \pm \frac{4 l^2}{l^2+1} dl d\tau
 - \frac{(l^2+1)^2}{4}dS^2
\label{ERotWHl}
\end{equation}

Really such a kind of metric usually is not considered as a ``true'' wormhole,
because it connects two different ``universes,'' but not domains of the same space. 
It should be mentioned also, that direction of ``proper time'' axis $\tau$ in 
second asymptotically flat region is opposite to the first one, it is denoted 
by "$\pm$" sign in \eq{ERotWH}. It either produces some difficulty for 
identification of two flat regions, or illustrate an idea about a close 
bond between wormholes and time machines \cite{WHWEC,BHTW}. Such kind of 
{\em acausal} processes also may be related with unitarity violation 
\cite{DeuTM,GottPres}, but discussion about the theory of wormholes, 
time machines, {\em etc.} is not a purpose of presented \paper.

\section{Conclusion}  

In present \paper\ is considered a possibility to write the Dirac equation 
covariant with respect to the group of general linear coordinate transformations.
It was not considered neither infinite-dimensional unitary representations 
nor methods of the quantum field theory, but even model of matrix Dirac
equation with finite-dimensional (16D) space was produced an interesting result. 
It was shown, that the matrix Dirac equation formally may be considered as an 
equation for two particles and the general linear coordinate transformation produce
specific ``entanglement'' unlike of Lorentz ones. 

The initial Dirac spinor may be considered as a formal subsystem and after general
linear coordinate transformation the state becomes entangled with second system,
{\em i.e.} for description it is necessary to use some analogue of 
{\em mixed state}. To fasten on a possible relation with the blackhole entropy, 
in last section was considered the Scwhartchild metric and it was shown,
that using of the general linear coordinate transformations may be 
appropriate for such a case.

It should be mentioned also, that currently the set of questions about relation 
of black hole thermodynamics, information theory and quantum mechanics in state 
of active development, {\em e.g.} already at the time of finishing of given 
\paper\ (March 2004) appear few fresh works with different approaches 
\cite{Wern,Terno,Yurts,Ng}.

\section*{Acknowledgments}
Author is grateful to A. Grib, D. Finkelstein, S. Krasnikov, Yu. Pavlov,
J. Bekenstein, D. Terno, I. Kirsch, E. Poberii, R. Zapatrin, B. Binder,
A. Lobashov and others.


\tableofcontents
\end{document}